\documentclass[12pt]{article}

\usepackage{amsmath}
\usepackage{amssymb}
\usepackage{graphicx}
\usepackage{caption2}
\usepackage{amsfonts}
\usepackage{cite}
\usepackage{lineno}
\usepackage{color}
\usepackage{multirow}
\usepackage{subfigure}
\usepackage{bm}

\newcommand{\be}{\begin{equation}}
\newcommand{\ee}{\end{equation}}
\newcommand{\bea}{\begin{eqnarray}}
\newcommand{\eea}{\end{eqnarray}}
\newcommand{\bef}{\begin{figure}}
\newcommand{\ef}{\end{figure}}
\newcommand{\bt}{\begin{tabular}}
\newcommand{\et}{\end{tabular}}
\newcommand{\bno}{\begin{enumerate}}
\newcommand{\eno}{\end{enumerate}}

\def\3{\ss}

\catcode`\"=\active
\def"{\accent'177}

\begin{document}

\begin{center}
{\bf\large  Spatial symmetry invariance of solution of  Kolmogorov flow}

Shijun Liao\footnote{Corresponding author, Email: sjliao@sjtu.edu.cn}\\

State Key Laboratory of Ocean Engineering, Shanghai, China\\
School of Ocean and Civil Engineering, Shanghai Jiao Tong University, China
\end{center}

\begin{abstract}
We prove a mathematical theorem that solution for all $t > 0$  of the Kolmogorov flow governed by Navier-Stokes (NS) equations  with periodic boundary condition keeps the same spatial symmetry as its initial condition.   The mathematical proof of a similar theorem for the three-dimensional incompressible NS equations  is given in the appendix.  These mathematical theorems can be used to check the correctness and reliability of numerical simulations of NS turbulence.   For example, they support the corresponding CNS (clean numerical simulation) results of the 2D and 3D turbulent Kolmogorov flows  \cite{Qin2024JOES, Liao-2025-JFM-NEC,Qin2025PoF} that remain the same spatial symmetry in the whole time interval of simulation, but do not support the corresponding DNS (direct numerical simulation) results that lose the spatial symmetry quickly.  In other words, these DNS results violate these mathematical theorems. Thus, these mathematical theorems rigorously confirm that the spatiotemporal trajectories of  NS turbulence given by DNS are indeed quickly polluted by numerical noises badly.  All of these also indicate that CNS indeed can provide helpful enlightenments to deepen our understanding about turbulence and besides approach some mathematical truths about the NS equations.       
\end{abstract}

\hspace{-0.4
cm}{\bf Keyword} Navier-Stokes equations, Kolmogorov flow,  spatial symmetry,  turbulence, CNS, DNS

\hspace{-0.4cm}{\bf MSC} 76F02, 76F65, 65M25

\section{Introduction}

Lorenz,  who put forward in 1963 the famous  ``butterfly-effect''  of chaos~\cite{lorenz1963deterministic}, further discovered in 2006 that  numerical errors could have significant influences on global properties of chaotic systems~\cite{Lorenz2006Tellus}, because unavoidable numerical noises as small artificial disturbance exponentially increase to macro-level.   Note that the Navier-Stokes (NS) turbulence (i.e. turbulence  governed by NS equations) is chaotic~\cite{Deissler1986PoF,  Vassilicos2023JFM,  boffetta2001predictability, berera2018chaotic, boffetta2017chaos}.  However, for all traditional numerical algorithms including the direct numerical simulation (DNS) of NS equations, numerical noises are unavoidable.  Thus, logically speaking, according to the butterfly-effect of chaotic system, numerical noises of DNS of NS turbulence  should quickly increase to a macro-level, which certainly should lead to a distinct departure of numerical trajectory from exact solution of NS equations.  Thus, logically speaking,  DNS of NS turbulence  should be quickly polluted by numerical noises badly.  

Liao~\cite{Liao2009} proposed the so-called ``clean numerical simulation'' (CNS) to overcome the restrictions of the traditional numerical algorithms for chaos and turbulence.   Unlike DNS, numerical noises of CNS can be rigorously negligible in a finite   time-interval, which is however long enough for statistics, by means of reducing both truncation error and  round-off error to a required small enough level~\cite{Liao2023book,Hu2020JCP, Liao2022AAMM, Liao2022NA, Qin2022JFM, Qin2024JOES, Liao-2025-JFM-NEC,Qin2025PoF}.  Thus, CNS can give very accurate spatiotemporal trajectory of NS equations in a finite but long enough interval of time:  they are very close to the exact solution.  

Note that the numerical noises of CNS are much smaller than those of DNS.  So, CNS results can be used as benchmark solution to check the correctness and reliability of DNS results. In fact, some comparisons were given for two-dimensional (2D) incompressible Kolmogorov flow~\cite{Qin2024JOES, Liao-2025-JFM-NEC}: the numerical simulations  given by CNS remain the same spatial symmetry (in a finite but long enough time-interval) as the corresponding initial condition, but the corresponding DNS results   remain the same spatial symmetry only at the beginning but quickly lose it.  Thus, spatial symmetry becomes a key point to check the reliability of numerical simulation of NS turbulence.        

In this paper, we prove a mathematical theorem that exact solution   of the 2D Kolmogorov flow considered in \cite{Qin2024JOES, Liao-2025-JFM-NEC} always remains the same spatial symmetry for all $t > 0$ as its initial condition.  Note that the CNS results in \cite{Qin2024JOES, Liao-2025-JFM-NEC} agree well with this theorem, but the corresponding DNS results lose the spatial symmetry and thus violate this mathematical theorem, clearly confirming that DNS results are indeed quickly polluted by numerical noises badly, and depart distinctly away from exact solution.  The detailed mathematical proofs are given in \S~2,  the concluding remarks and discussions  are given in \S~3.

\section{Spatial symmetry invariance of solution}
           
Consider a 2D incompressible Kolmogorov flow \cite{obukhov1983kolmogorov, chandler2013invariant, wu2021quadratic} in a square domain under the so-called Kolmogorov forcing, governed by the  dimensionless Navier-Stokes equation in the form of stream-function 
\begin{equation}
 \frac{\partial}{\partial t}\Big(\nabla^{2}\psi\Big)
 +\psi_x\nabla^{2}\psi_y-\psi_y\nabla^{2}\psi_x
 -\frac{1}{Re}\nabla^{4}\psi+n_K\cos(n_Ky)=0,       \label{eq_psi}
\end{equation}
where $\psi$ is the stream-function,  $t$ denotes the time, $x,y\in[0,2\pi]$ are horizontal and vertical coordinates in the coordinate system $O-xy$,  $Re$ is the Reynolds number,   $n_K$ is an even number describing the Kolmogorov forcing scale,  
  $\nabla^{2}$ is the Laplace operator,  and  $\nabla^{4}=\nabla^{2}\nabla^{2}$,  respectively. The stream-function $\psi$ satisfies the periodic boundary condition
\begin{equation}
\psi(x, y, t)=\psi(x\pm2\pi, y, t)=\psi(x, y\pm2\pi, t).       \label{boundary_condition}
\end{equation}
Here, let us consider a smooth initial condition $\psi(x,y,0)$ that contains the spatial symmetry of rotation
\begin{equation}
\psi(x,y,0)=\psi(2\pi-x,2\pi-y,0)   \label{symmetry_psi:A}
\end{equation}
and/or translation
\begin{equation}
 \psi(x,y,0)=\psi(x+\pi,y+\pi,0).         \label{symmetry_psi:B}
\end{equation}
        
  Define      
 \begin{equation}
  \psi^{(m)}(x,y,t) =  \frac{1}{m!} \frac{\partial^m \psi(x,y,t)}{\partial t^m}
 \end{equation}
 with 
 \begin{equation}
  \psi^{(0)}(x,y,t) =  \psi(x,y,t).   
 \end{equation}
Assume that $\psi(x,y,t)$ can be expanded as Taylor series at $t=t_{0}\in [0,+\infty)$ with a non-zero radius $\rho>0$ of convergence, say,
 \begin{eqnarray}
 \psi(x,y,t) &=& \psi(x,y,t_0)+\sum_{m=1}^{+\infty} \psi^{(m)}(x,y,t_0)(t-t_{0})^m \nonumber \\ 
 &=&\sum_{m=0}^{+\infty} \psi^{(m)}(x,y,t_0)(t-t_{0})^m,
 \end{eqnarray}
 where $|t-t_{0}| < \rho$.  Then, we can rewrite (\ref{eq_psi}) as follows  
\begin{equation}
\nabla^{2}\psi^{(1)}(x,y,t) 
 +\psi^{(0)}_x\nabla^{2}\psi^{(0)}_y-\psi^{(0)}_y\nabla^{2}\psi^{(0)}_x
 -\frac{1}{Re}\nabla^{4}\psi^{(0)}+n_K\cos(n_Ky)=0.       \label{eq_psi:A}
\end{equation}
 
Let us consider another coordinate system $O'-x'y'$, where  
\begin{equation}
x' = 2\pi-x, \;\;  y' = 2\pi-y \label{x-y:rotation}
\end{equation}
for rotation, and
\begin{equation}
x' = \pi+x, \;\;  y' = \pi+y \label{x-y:translation}
\end{equation}
for translation, respectively, in which the corresponding solution reads  $\psi'(x',y',t)$.   
Since $\psi'(x',y',t) $ satisfies the same equations, it holds 
\begin{equation}
\nabla'^{2}\psi'^{(1)} 
 +\psi'_{x'}\nabla'^{2}\psi'_{y'}-\psi'_{y'}\nabla'^{2}\psi'_{x'}
 -\frac{1}{Re}\nabla'^{4}\psi'+n_K\cos(n_K y')=0       \label{eq_psi:B}
\end{equation}
under the periodic boundary condition, subject to an initial condition $\psi'(x',y',0)$.  
 Since there exists the spatial symmetry of rotation~(\ref{symmetry_psi:A}) and/or translation~(\ref{symmetry_psi:B}) for the initial solution, it holds 
\begin{equation}
 \psi'(x',y',0) = \psi(x,y,0).   
\end{equation}

\begin{description}
\item[Lemma 1]  {\em Let $n_{K}\geq 2$ denote an even integer, $f$ and $g$ denote two smooth functions with spatial symmetry $f(x,y,t)=f'(x',y',t)$ and $g(x,y,t)=g'(x',y',t)$, respectively.   It holds  
\begin{eqnarray}
\cos(n_{k} y) =\cos(n_{k}y'), \;\;\; \nabla^{2} =  \nabla'^{2},\;\;\; \nabla^{4} = \nabla'^{4}, \;\;\;
 \frac{\partial f}{\partial x} \; \frac{\partial g}{\partial y} = \frac{\partial f'}{\partial x'} \; \frac{\partial g'}{\partial y'} 
 \label{res:Lemma1}
\end{eqnarray}
 in the case of (\ref{x-y:rotation}) and/or (\ref{x-y:translation}).}   
\end{description}
 Proof. In the case of (\ref{x-y:rotation}),  we have 
 \begin{equation}
\cos(n_K y') =  \cos(2 n_K \pi  - n_k y) = \cos(n_K y),  \label{cos:A}
\end{equation} 
 \begin{eqnarray}
\frac{\partial }{\partial x} = \frac{\partial }{\partial x'} \frac{\partial x'}{\partial x} =- \frac{\partial }{\partial x'}, \;\;\; 
\frac{\partial }{\partial y} = \frac{\partial }{\partial y'} \frac{\partial y'}{\partial y} =- \frac{\partial }{\partial y'},  \label{derivative:A}
\end{eqnarray}
\begin{eqnarray}
\frac{\partial^2 }{\partial x^2} =  - \frac{\partial }{\partial x'} \left( - \frac{\partial }{\partial x'} \right)  =  \frac{\partial^2 }{\partial x'^2},  \;\;
\frac{\partial^2 }{\partial y^2} =  - \frac{\partial }{\partial y'} \left( - \frac{\partial }{\partial y'} \right)  =  \frac{\partial^2 }{\partial y'^2}, 
\end{eqnarray}
so that 
\begin{eqnarray}
\nabla^{2}  &=&  \frac{\partial^2 }{\partial x^2} +  \frac{\partial^2 }{\partial y^2}  =  \frac{\partial^2 }{\partial x'^2} +  \frac{\partial^2 }{\partial y'^2} = \nabla'^{2}, \label{nabla:2:A}
\end{eqnarray}
\begin{eqnarray}
\nabla^{4} =  \nabla^{2}\; \nabla^{2} = \nabla'^{2}\; \nabla'^{2} = \nabla'^{4},  \label{nabla:4:A}
\end{eqnarray}
and 
\begin{eqnarray}
\frac{\partial f}{\partial x} \; \frac{\partial g}{\partial y} = \left(- \frac{\partial f'}{\partial x'} \right)  \left(- \frac{\partial g'}{\partial y'} \right) = \frac{\partial f'}{\partial x'} \; \frac{\partial g'}{\partial y'} 
\end{eqnarray}
for  two smooth functions $f,g$ with the spatial symmetry $f=f'$ and $g=g'$.  

Similarly, in the case of (\ref{x-y:translation}), we have
\begin{equation}
\cos(n_K y') =  \cos( n_K \pi + n_k y) = \cos(n_K y),  \label{cos:B}
\end{equation}
since $n_k$ is an even number, and   
\begin{eqnarray}
\frac{\partial }{\partial x} = \frac{\partial }{\partial x'}, \;\;\; 
\frac{\partial }{\partial y}  =  \frac{\partial }{\partial y'}, \;\;\; 
\frac{\partial^2 }{\partial x^2} = \frac{\partial^2 }{\partial x'^2}, \;\;\; 
\frac{\partial^2 }{\partial y^2} = \frac{\partial^2 }{\partial y'^2},   \label{derivative:B}
\end{eqnarray}
so that 
\begin{equation}
\nabla^{2}=   \nabla'^{2},  \nabla^{4} =   \nabla'^{4} 
\end{equation}
and
\begin{eqnarray}
\frac{\partial f}{\partial x} \; \frac{\partial g}{\partial y} = \frac{\partial f'}{\partial x'} \; \frac{\partial g'}{\partial y'} 
\end{eqnarray}
for two smooth functions $f,  g$ with the spatial symmetry $f=f'$ and $g=g'$.  This ends the proof of Lemma~1.

\begin{description}
\item[Lemma 2]  {\em For the two-dimensional Kolmogorov flow governed by the Navier-Stokes equation (\ref{eq_psi}) (with an even number $n_{K}$) under the periodic boundary condition (\ref{boundary_condition}) subject to a smooth initial condition with the spatial symmetry of rotation (\ref{symmetry_psi:A}) and/or translation (\ref{symmetry_psi:B}), if its solution $\psi(x,y,t)$ at $t=t_{0}\geq 0$ contains a spatial symmetry $\psi(x,y,t_{0})=\psi'(x',y',t_{0})$ of rotation and/or translation, then there exists the spatial symmetry $\psi^{(m)}(x,y,t_{0}) = \psi'^{(m)}(x',y',t_{0})$ for all integers $m\geq 1$. }
\end{description}
Proof. This Lemma can be proved by means of the recursive method.  

\hspace{-0.75cm}{(A) When $m=1$}

Since there exists the spatial symmetry at $t=t_0$, it holds
\begin{equation}
\psi'(x',y',t_0) = \psi(x,y,t_0),  \mbox{i.e.} \hspace{0.3cm} \psi'^{(0)}(x',y',t_0) = \psi^{(0)}(x,y,t_0).    \label{symmetry:0}
\end{equation}

According to (\ref{res:Lemma1}) of Lemma~1 and (\ref{symmetry:0}), we have at $t=t_{0}$ that 
\begin{eqnarray}
&&\psi'^{(0)}_{x'}\nabla'^{2}\psi'^{(0)}_{y'}-\psi'^{(0)}_{y'}\nabla'^{2}\psi'^{(0)}_{x'}
 -\frac{1}{Re}\nabla'^{4}\psi'^{(0)}  \nonumber \\
 &=&\psi'^{(0)}_{x'} \left(\nabla'^{2}\psi'^{(0)}\right)_{y'} - \psi'^{(0)}_{y'} \left( \nabla'^{2}\psi'^{(0)}\right)_{x'}
 -\frac{1}{Re}\nabla^{4}\psi^{(0)}  \nonumber \\
   &=&\psi^{(0)}_{x} \left(\nabla^{2}\psi^{(0)}\right)_{y} - \psi^{(0)}_{y} \left( \nabla^{2}\psi^{(0)}\right)_{x}
 -\frac{1}{Re}\nabla^{4}\psi^{(0)}   \nonumber \\
 &=&\psi^{(0)}_{x}\nabla^{2}\psi^{(0)}_{y}-\psi^{(0)}_{y}\nabla^{2}\psi^{(0)}_{x}
 -\frac{1}{Re}\nabla^{4}\psi^{(0)}, 
\end{eqnarray}
so that (\ref{eq_psi:B}) becomes 
\begin{equation}
\nabla^{2}\psi'^{(1)}
 +\psi^{(0)}_{x}\nabla^{2}\psi^{(0)}_{y}-\psi^{(0)}_{y}\nabla^{2}\psi^{(0)}_{x}
 -\frac{1}{Re}\nabla^{4}\psi^{(0)} +n_K\cos(n_K y)=0.       \label{eq_psi:A:B}
\end{equation}
Comparing (\ref{eq_psi:A:B}) to  (\ref{eq_psi:A})  gives the spatial symmetry at $t=t_{0}$ that
\begin{equation}
\psi'^{(1)}(x',y',t_{0}) = \psi^{(1)}(x,y,t_{0}).  \label{symmetry:1}
\end{equation}
Thus, Lemma~2 holds true when $m=1$. 

\hspace{-0.75cm}(B) Assume that Lemma~{2} holds true when $m=1,2,\cdots, n$, say,
 there exists the spatial symmetry at $t =t_{0}$:
\begin{equation}
\psi'^{(i)}(x',y',t_{0}) = \psi^{(i)}(x,y,t_{0}), \hspace{1.0cm}  0\leq i \leq n.   \label{symmetry:m}
\end{equation}

Differentiating  (\ref{eq_psi}) and  (\ref{eq_psi:B}) $n$ times  with respect to $t$ and then dividing by $n!$, we have 
\begin{eqnarray}
(n+1) \nabla^2 \psi^{(n+1)} 
= \frac{1}{Re}\nabla^{4}\psi^{(n)}- \sum_{i=0}^{n}\Big[ \psi^{(i)}_{x}\nabla^{2}\psi^{(n-i)}_{y}-\psi^{(i)}_{y}\nabla^{2}\psi^{(n-i)}_{x}
 \Big] \label{eqn:psi:m:A}  
\end{eqnarray}
in the coordinate system $O-xy$, and  
\begin{eqnarray}
(n+1) \nabla'^2 \psi'^{(n+1)} 
= \frac{1}{Re}\nabla'^{4}\psi'^{(n)}-  \sum_{i=0}^{n}\left[ \psi'^{(i)}_{x'}\nabla'^{2}\psi'^{(n-i)}_{y'}-\psi'^{(i)}_{y'}\nabla'^{2}\psi'^{(n-i)}_{x'}
 \right]  \label{eqn:psi:m:B}  
\end{eqnarray}
in the coordinate system $O'-x'y'$, respectively.  

According to (\ref{res:Lemma1}) of Lemma~1 and (\ref{symmetry:m}), we have at $t=t_{0}$ that 
\begin{equation}
\nabla'^2 \psi'^{(n+1)}=\nabla^2 \psi'^{(n+1)},\;\;\; \nabla'^{4}\psi'^{(n)}= \nabla^{4}\psi^{(n)}, \label{nabla:m:B}
\end{equation}
and
\begin{eqnarray}
&& \sum_{i=0}^{n}\left[ \psi'^{(i)}_{x'}\nabla'^{2}\psi'^{(n-i)}_{y'}-\psi'^{(i)}_{y'}\nabla'^{2}\psi'^{(n-i)}_{x'}
 \right]  \nonumber\\
 &=&  \sum_{i=0}^{n}\left[ \psi'^{(i)}_{x'}\left( \nabla'^{2}\psi'^{(n-i)}\right)_{y'}-\psi'^{(i)}_{y'}\left(\nabla'^{2}\psi'^{(n-i)}\right)_{x'}
 \right]  \nonumber\\
 &=&  \sum_{i=0}^{n}\left[ \psi^{(i)}_{x}\left( \nabla^{2}\psi^{(n-i)}\right)_{y}-\psi^{(i)}_{y}\left(\nabla^{2}\psi^{(n-i)}\right)_{x}
 \right]  \nonumber\\
 &=& \sum_{i=0}^{n}\Big[ \psi^{(i)}_{x}\nabla^{2}\psi^{(n-i)}_{y}-\psi^{(i)}_{y}\nabla^{2}\psi^{(n-i)}_{x}
 \Big].  \label{eqn:F:m}
\end{eqnarray}

Substituting  (\ref{nabla:m:B}) and (\ref{eqn:F:m})  into (\ref{eqn:psi:m:B}) gives 
\begin{eqnarray}
(n+1) \nabla^2 \psi'^{(n+1)}
  = \frac{1}{Re}\nabla^{4}\psi^{(n)}- \sum_{i=0}^{n}\Big[ \psi^{(i)}_{x}\nabla^{2}\psi^{(n-i)}_{y}-\psi^{(i)}_{y}\nabla^{2}\psi^{(n-i)}_{x} \Big].  \label{eqn:psi:m:A:B}  
\end{eqnarray}
Comparing (\ref{eqn:psi:m:A:B}) to (\ref{eqn:psi:m:A}) gives the spatial symmetry 
\begin{equation}
\psi'^{(n+1)}(x',y',t_{0}) = \psi^{(n+1)}(x,y,t_{0})
\end{equation}
for  integer $n\geq 1$.  Thus, if Lemma~2 holds for an integer $n\geq 1$, it also holds true for the integer $n+1$.

\hspace{-0.75cm}{(C)}  Based on (A) and (B), we have due to the the recursive method that 
 the spatial symmetry 
\begin{equation}
\psi'^{(m)}(x',y',t_{0}) = \psi^{(m)}(x,y,t_{0}),  \label{symmetry:arbitrary:m}
\end{equation}
holds for an arbitrary integer $m\geq 1$.  This ends the proof of Lemma~2. 

\begin{description}
\item[Lemma 3] 
{\em For the two-dimensional Kolmogorov flow governed by the Navier-Stokes equation (\ref{eq_psi}) (with an even number $n_{K}$) under the periodic boundary condition (\ref{boundary_condition}) subject to a smooth initial condition with the spatial symmetry of rotation (\ref{symmetry_psi:A}) and/or translation (\ref{symmetry_psi:B}), if its solution $\psi(x,y,t)$  contains a spatial symmetry at $t=t_{0}$, say, $\psi'(x',y',t_{0})=\psi(x,y,t_{0})$, and besides  its Taylor series exists at $t=t_{0}$  with a non-zero  radius $\rho$  of convergence, then $\psi(x,y,t)$ in  $t\in[t_{0}, t_{0}+\rho)$ has the same spatial symmetry as  $\psi(x,y,t_{0})$.     }
\end{description}

 \hspace{-0.75cm}Proof:  According to Lemma~2, from the spatial symmetry $\psi'(x',y',t_{0})=\psi(x,y,t_{0})$,  we have the spatial symmetry 
\begin{equation}
\psi'^{(m)}(x',y',t_{0}) = \psi^{(m)}(x,y,t_{0}), \hspace{1.0cm}  1\leq m <+\infty.   
\end{equation}
Then, 
 \begin{eqnarray}
&& \psi'(x',y',t) \nonumber\\
&=& \sum_{m=0}^{+\infty} \psi'^{(m)}(x',y',t_0)\; (t-t_{0})^m = \sum_{m=0}^{+\infty} \psi^{(m)}(x,y,t_0) (t-t_{0})^m \nonumber\\
&=& \psi(x,y,t), 
 \end{eqnarray}
where $|t-t_{0}| < \rho$, including $t \in [t_{0}, t_{0}+\rho)$, where $\rho>0$ is the radius of convergence.  This guarantees  the spatial symmetry $\psi'(x',y',t)=\psi(x,y,t)$ in $t \in [t_{0}, t_{0}+\rho)$.    This ends the proof of Lemma~3.       

\begin{description}
\item[Theorem of spatial symmetry invariance] {\em 
For the 2D Kolmogorov flow governed by the Navier-Stokes equation (\ref{eq_psi}) (with an even number $n_{K}$) under the periodic boundary condition (\ref{boundary_condition}) subject to a smooth initial condition with the spatial symmetry of rotation (\ref{symmetry_psi:A}) and/or translation (\ref{symmetry_psi:B}), if its Taylor series exists with a non-zero  convergence radius $\rho$ for all $t\geq 0$, its solution for all $t>0$ keeps the same spatial symmetry as  its initial condition.   }
\end{description}

 \hspace{-0.75cm}Proof: Since the initial condition contains the spatial symmetry, i.e. $\psi(x,y,0)=\psi'(x',y',0)$, we have according to Lemma~3 (by setting $t_{0} = 0$) the spatial symmetry  
\[  \psi(x, y, t)=\psi'(x', y', t), \hspace{1.0cm} t\in[0,\delta_{0}], \]
where $\delta_{0}<\rho_{0}$, and $\rho_{0}$ is the radius of convergence of its Taylor series at $t=0$.  Then, setting $t_{0}=\delta_{0}$, we have according to  Lemma~3 the spatial symmetry 
\[  \psi(x, y, t)=\psi'(x', y', t), \hspace{1.0cm} t\in[\delta_{0},\delta_{1}], \]
where $\delta_{1}<\rho_{1}$, and $\rho_{1}$ is the radius of convergence of its Taylor series at $t=t_{0}=\delta_{0}$.
Furthermore, setting $t_{0}=\delta_{1}$, we have according to  Lemma~3 the spatial symmetry  
\[  \psi(x, y, t)=\psi'(x', y', t), \hspace{1.0cm} t\in[\delta_{1},\delta_{2}], \]
where  $\delta_{2}<\rho_{2}$, and $\rho_{2}$ is the radius of convergence of its Taylor series at $t=t_{0}=\delta_{1}$. 
Note that 
\[   [0,\delta_{0}] \cup [\delta_{0},\delta_{1}] \cup [\delta_{1},\delta_{2}] = [0,\delta_{2}],  \]
thus now there exists the spatial symmetry 
\[  \psi(x, y, t)=\psi'(x', y', t), \hspace{1.0cm} t\in[0,\delta_{2}]  \]
in a larger interval of time. The same process can keep going, since the Taylor series of $\psi(x,y,t)$ always exists with a non-zero radius of convergence.  
Thus,  it always holds the spatial symmetry 
\[  \psi(x, y,  t)=\psi'(x', y', t), \;\; t\in[0,+\infty). \] 
This ends the proof of the theorem.

\section{Remarks and discussions}

 \begin{description}
\item[Remark 1]     Let $O$ denote one observer in the coordinate system $O-xy$, and $O'$ denote another observer in the coordinate system $O'-x'y'$, respectively, where $(x',y')=(x+\pi, y+\pi)$ and/or  $(x',y')=(2\pi-x, 2\pi-y)$.   When there exist the spatial symmetry  $\psi'(x',y',0) = \psi(x,y,0)$ for the initial condition,  the governing equations and the initial condition are the {\em same} for both of the observer $O$ in the coordinate system $O-xy$ and the observer $O'$ in the  coordinate system  $O'-x'y'$.  Thus, physically speaking, the two observers at the two different coordinate systems should observe the same flow.  From the viewpoint of the third observer simultaneously knowing the relationship between their coordinate systems and their results, the flow governed by NS equations  (\ref{eq_psi}) should keep the same spatial symmetry as its initial condition.  This physical insight is indeed true,  as mathematically proved  in this paper.  So,   the theorem of spatial symmetry invariance is easy to understand from physical viewpoint.  
\end{description}

\begin{description}
\item[Remark 2]  Numerical simulations of the 2D Kolmogorov flow governed by the Navier-Stokes equation (\ref{eq_psi}) (with an even number $n_{K}$) under the periodic boundary condition (\ref{boundary_condition}) subject to a smooth initial condition with the spatial symmetry of rotation (\ref{symmetry_psi:A}) and/or translation (\ref{symmetry_psi:B}) distinctly deviate exact solution, if they violate the theorem of spatial symmetry invariance proved in this paper.   Thus, this theorem can be used to check the correctness and accuracy of numerical simulations of 2D turbulent Kolmogorov flow.   
\end{description}

For example, 
as reported in \cite{Qin2024JOES, Liao-2025-JFM-NEC}, the numerical simulations of the 2D turbulent Kolmogorov flow given by CNS (in a finite but long enough time-interval) always remain the same spatial symmetry as the corresponding initial condition.   This agrees well with the theorem of spatial symmetry  invariance.   This is mainly because, unlike traditional algorithms, numerical noises of CNS are rigorously negligible in a finite time-interval that is however long enough for statistics.    
However, as mentioned in \cite{Qin2024JOES, Liao-2025-JFM-NEC},  the corresponding DNS results remain the same spatial symmetry as the corresponding initial condition only in a short duration from the beginning but then quickly lose the spatial symmetry, clearly indicating that DNS results distinctly deviate the exact solution of the NS equations.  In other words, these DNS results violate  the theorem of spatial symmetry invariance.  This clearly indicates that the DNS results are indeed  quickly polluted by numerical noises badly.  
The above conclusions  should have general meanings for turbulent flows, as illustrated in \cite{Qin2022JFM}.

\begin{description}
\item[Remark 3]  For the 2D Kolmogorov flow governed by the Navier-Stokes equation (\ref{eq_psi}) (with an even number $n_{K}$) under the periodic boundary condition (\ref{boundary_condition}) subject to a smooth initial condition with the spatial symmetry of rotation (\ref{symmetry_psi:A}) and/or translation (\ref{symmetry_psi:B}), if initial condition has different spatial symmetries,  their corresponding solutions should be distinctly different even in statistics.     
\end{description}

For example, Liao and Qin~\cite{Liao-2025-JFM-NEC} applied CNS to solve the 2D turbulent Kolmogorov flow (\ref{eq_psi}) in the case of   
$Re=2000$ and $n_{K} = 16$, subject to the different initial conditions
\begin{eqnarray}
\psi_{1}(x,y,0) &=& -\frac{1}{2}\left[ \cos(x+y) + \cos(x-y)\right],\label{IC:1}\\
 \psi_{2}(x,y,0) &=& -\frac{1}{2}\left[ \cos(x+y) + \cos(x-y)\right] + \delta' \; \sin(x+y), \label{IC:2}
\end{eqnarray}  
where $\delta'$ is a constant.  Note that $\psi_{1}(x,y,0)$ contains the spatial symmetry of both rotation and translation, but  $\psi_{2}(x,y,0)$ contains the spatial symmetry of only translation for a large $\delta' \sim O(1)$.   As reported by Liao and Qin~\cite{Liao-2025-JFM-NEC}, the spatiotemporal trajectories given by CNS using the initial condition $\psi_{1}(x,y,0)$ remains the same spatial symmetry of rotation and translation as the initial condition  $\psi_{1}(x,y,0)$ in the whole time-interval $t\in[0,300]$.  This agrees well with the theorem of spatial symmetry invariance proved in~\S~2.  When $\delta'=10^{{-20}}$ in (\ref{IC:2}), the spatiotemporal trajectories given by CNS using the initial condition $\psi_{2}(x,y,0)$ seems to remain the same spatial symmetry of rotation and translation as the initial condition $\psi_{1}(x,y,0)$ at the beginning, mainly because the disturbance $10^{-20} \sin(x+y)$ of $\psi_{2}(x,y,0)$ is too small compared to $\psi_{1}(x,y,0)$, until  it enlarges to a macro-level,  due to the so-called ``butterfly-effect'' of the chaotic characteristics of turbulence~\cite{Deissler1986PoF,  Vassilicos2023JFM,  boffetta2001predictability, berera2018chaotic, boffetta2017chaos}, at about $t\approx 35$ when it finally destroys the original spatial symmetry of $\psi_{1}(x,y,0)$, thus the CNS result thereafter remains the same spatial symmetry of translation as its initial condition $\psi_{2}(x,y,0)$ for large $\delta' \sim O(1)$, as reported by Liao and Qin~\cite{Liao-2025-JFM-NEC}.  This CNS result  also agrees well with the theorem of spatial symmetry  invariance.   Certainly, the two CNS results should have different statistics, since they have different spatial symmetries.  So, the theorem of spatial symmetry invariance also supports the conclusion that the turbulence is a kind of ultra-chaos~\cite{Liao2022AAMM,qin2025ultrachao-NS-arXiv}, i.e. small disturbance could lead to distinct  differences not only in spatotemporal trajectory but also in statistics of turbulence.   
It should be emphasized again that, the corresponding DNS results of the same turbulent Kolmogorov flows quickly lose their spatial symmetry, since random numerical noises without spatial symmetry quickly enlarges to a macro-level.  Therefore, the theorem of spatial symmetry invariance  rigorously confirms that the DNS results of the 2D turbulent Kolmogorov flow (\ref{eq_psi}) are indeed quickly polluted by numerical noises badly, as illustrated in \cite{Qin2024JOES}.   This is also the reason why the so-called ``noise-expansion cascade'' revealed in~\cite{Liao-2025-JFM-NEC} by means of CNS cannot be discovered by means of DNS.  

For  a large enough Reynolds number such as $Re=2000$ considered in \cite{Liao-2025-JFM-NEC}, the 2D Kolmogorov flow (\ref{eq_psi}) is turbulent.   It is widely accepted by fluid community that turbulence is chaotic~\cite{Deissler1986PoF,  Vassilicos2023JFM,  boffetta2001predictability, berera2018chaotic, boffetta2017chaos} and thus, due to the famous butterfly-effect of chaotic system,  the small disturbance $\delta' \sin(x+y)$ of the initial condition $\psi_{2}(x,y,t)$ defined  by (\ref{IC:2}) should be quickly enlarged to the same order of magnitude as the first initial condition  $\psi_{1}(x,y,t)$ described by (\ref{IC:1}).  However,  $\psi_{1}(x,y,t)$ has a different spatial symmetry from that of  $\psi_{2}(x,y,t)$ when $\delta' \sim O(1)$, therefore, according to the theorem of spatial symmetry invariance, the turbulent Kolmogorov flow  $\psi_{1}(x,y,t)$ has different spatial symmetry from that of  $\psi_{2}(x,y,t)$ even for arbitrarily small $\delta'$, say, $\delta' \to 0$.  Thus, the theorem of spatial symmetry invariance plus the butterfly-effect of chaos  supports  the following conjecture:
\begin{description}
\item[Conjecture A] {\em The two-dimensional turbulent Kolmogorov flow governed by the NS equation (\ref{eq_psi}) (with an even number $n_{K}$) under the periodic boundary condition (\ref{boundary_condition}) subject to a smooth initial condition with the spatial symmetry of rotation (\ref{symmetry_psi:A}) and/or translation (\ref{symmetry_psi:B}) admit distinct smooth global solutions for all $t > 0$ from almost the same initial conditions $\psi_{1}(x,y,0)$ and $\psi_{2}(x,y,0)$ with $\big|\psi_{1}(x,y,0)-\psi_{2}(x,y,0)\big| \leq \delta$ for arbitrarily small $\delta$ (or as $\delta \to 0$). }
\end{description}

As reported by Qin and Liao~\cite{Qin2025PoF},  the CNS results of the three-dimensional (3D) Kolmogorov flow also remain the same spatial symmetry as its initial condition in a finite but long enough interval of time $t\in[0,500]$.  This highly suggests that  Conjecture A might have general meanings.  All of these support   
 the following conjecture mentioned by Liao in \cite{Liao2026nonuniqueness-arXiv}:
\begin{description}
\item[Conjecture B] {\em The Navier-Stokes equations admit distinct smooth global solutions ${\bf u}_{1}({\bf r},t)$ and ${\bf u}_{2}({\bf r},t)$ for all $t > 0$ from almost the same initial conditions ${\bf u}_{1}({\bf r},0)$ and ${\bf u}_{2}({\bf r},0)$ with $\big|{\bf u}_{1}({\bf r},0)-{\bf u}_{2}({\bf r},0)\big| \leq \delta$ for arbitrarily small $\delta$ (or as $\delta \to 0$). }
\end{description}
Obviously, if the flow is laminar, all small disturbances in the initial condition should not increase to a macro-level, say, remain small forever.   Thus, the above two conjectures should hold true only for turbulent flows.    

What happens when the initial condition has no spatial symmetry?  Qin and Liao~\cite{Qin2022JFM} applied both of DNS and CNS to solve the 2D Rayleigh-B\'{e}nard turbulent flow using thermal fluctuation as initial condition, which is ransom in space and thus has no spatial symmetry: it was found~\cite{Qin2022JFM}  that the flow given by CNS remains a kind of  vortical flow in the whole duration $t\in[0,500]$, but the flow given by DNS is first a  kind of vortical flow, until at $t\approx 188$ when it transfers into a kind of zonal flow thereafter in $t\in[188,500]$.   Currently, it was found that   the final flow type (given by the DNS)  of the 2D Rayleigh-B\'{e}nard turbulent flow using random thermal fluctuation as initial condition are very sensitive to the time-step for some Reynolds numbers~\cite{Qin2026paradox-arXiv}: the two flow types frequently alternate as the time-step is reduced to a very small value, suggesting that the time-step corresponding to each turbulent flow type should be densely distributed.  All of  these examples clearly illustrate that numerical noises indeed might have huge influences on global characteristics of turbulence, although sometimes.  It is an open question whether or not random initial condition without spatial symmetry {\bf must} be used for NS turbulence.  
   
Note that, unlike the {\bf exact} solution mentioned  in the theorem of spatial symmetry invariance proved in \S~2,  {\bf numerical}  results given by CNS are reliable only in a finite but long enough interval of time $t\in[0,T_{c}]$,  where $T_{c}$ is called  the ``critical predictable time''.  As pointed out by Liao in his book~\cite{Liao2023book}, 
\begin{equation} 
T_{c}\sim -\kappa \ln {\cal E}_{0}
\end{equation}
for chaotic system and turbulence, where ${\cal E}_{0}>0$ denotes the level of background numerical noises, and $\kappa>0$ is the Lyapunov exponent of chaotic system.  Obviously, the smaller the background numerical noise ${\cal E}_{0}$, the larger the so-called ``critical predictable time'' $T_{c}$.  This is the reason why CNS can give reliable simulations of chaotic systems and turbulent flows  in a much larger time-interval  $t\in[0,T_{c}]$ than DNS, because ${\cal E}_{0}$ of CNS is much smaller than that of DNS.  
From this viewpoint, we can regard DNS as a special case of CNS, because DNS often uses single or double precision with the 4th-order Runge-Kutta's method  in time dimension, corresponding to a rather small ``critical predictable time''  $T_{c}$ that is too short for statistic computation, although DNS has not  the  concept of ``critical predictable time''.  

It should be emphasized that, although CNS results are numerical, they can provide us enlightenments to approach some mathematical truths such as the theorem of spatial symmetry invariance proved in \S~2 and the conjectures  mentioned above and in \cite{Liao2026nonuniqueness-arXiv}.  Hopefully, CNS~\cite{Liao2009, Liao2023book,Hu2020JCP, Liao2022AAMM, Liao2022NA, Qin2022JFM, Qin2024JOES, Liao-2025-JFM-NEC,Qin2025PoF, qin2025ultrachao-NS-arXiv,Liao2026nonuniqueness-arXiv} could be used as a powerful tool to study turbulence.    

Finally,  the theorem of spatial symmetry invariance for 2D Kolmogorov flow can be generalized to the  three-dimensional (3D) incompressible Navier-Stokes equations~\cite{Batchelor2000}, as proved below  in Appendix in a similar way.  

\setcounter{equation}{0}

\renewcommand{\theequation}{A.\arabic{equation}}

\begin{center}

{\bf\Large  Appendix}

{\bf\Large Spatial symmetry invariance  of solution of the Navier-Stokes equations}

\end{center}

Let us consider the dimensionless  three-dimensional (3D) incompressible Navier-Stokes equations~\cite{Batchelor2000}: 
\begin{eqnarray}
\nabla \cdot {\bf u} &=& 0,  \label{eq:continuation} \\
 \frac{\partial {\bf u}}{\partial t}
 +\left( {\bf u}\cdot \nabla\right) {\bf u} &=&-\nabla p + \frac{1}{Re} \Delta {\bf u} + {\bf f},       \label{eq:u}
\end{eqnarray}
under periodic boundary condition  in a square domain ${\bf x}=(x_1,x_2,x_3)\in[0,2\pi]^3$,   
where $t\in[0,+\infty)$ denotes the time, ${\bf u}=(u_1,u_2,u_3)$ is the velocity of fluid,  $p$ is the pressure, ${\bf f}({\bf x})$ is an steady-state external force,  $Re$ is the Reynolds number,    $\nabla$ is the Hamilton operator,  and $\Delta=\nabla\cdot \nabla$ is the Laplace operator,  respectively.  Let $O-x_{1}x_{2}x_{3}$ denote its corresponding coordinate system.   
  
Define 
\begin{equation}
 {\bf u}^{(m)}({\bf x},t) =\frac{1}{m!} \frac{\partial {\bf u}({\bf x},t)}{\partial t^{m}},\;\;\;  p^{(m)}({\bf x},t) =\frac{1}{m!}  \frac{\partial p({\bf x},t)}{\partial t^{m}}, \;\;\; m\geq 0, \label{def:u[m]:p[m]}
 \end{equation}
 with 
 \begin{equation}
  {\bf u}^{(0)}({\bf x},t) =  {\bf u}({\bf x},t), \;\;\;  p^{(0)}({\bf x},t) = p({\bf x},t),
 \end{equation}
for integer $m\geq 0$.  

From (\ref{eq:u}), it holds 
\begin{eqnarray}
 \frac{\partial (\nabla\cdot {\bf u})}{\partial t}
 +\nabla\cdot  \left[\left( {\bf u}\cdot \nabla\right) {\bf u}\right] &=&-\Delta p + \frac{1}{Re} \Delta \left(\nabla\cdot {\bf u}\right) + \nabla\cdot {\bf f},    
\end{eqnarray}
which gives using the continuation equation (\ref{eq:continuation}) that
\begin{equation}
\Delta p =\nabla\cdot {\bf f} -\nabla\cdot  \left[\left( {\bf u}\cdot \nabla\right) {\bf u}\right]. \label{eq:p}    
\end{equation}
Differentiating (\ref{eq:u}) and (\ref{eq:p}) $m$ times with respect to $t$ and then dividing by $m!$, we have
\begin{equation}
(m+1){\bf u}^{(m+1)} 
 +\sum_{i=0}^{m}\left( {\bf u}^{(i)}\cdot \nabla\right) {\bf u}^{(m-i)} =-\nabla p^{(m)} + \frac{1}{Re} \Delta {\bf u}^{(m)} ,       \;\;  \label{eq:u:m}
\end{equation}
\begin{equation}
\Delta p^{(m)} =-\sum_{i=0}^{m}\nabla\cdot  \left[\left( {\bf u}^{(i)}\cdot \nabla\right) {\bf u}^{(m-i)}\right], \label{eq:p:m}    
\end{equation}
for $m\geq 1$ and $t\geq 0$. 

Let ${\bf u}'({\bf x}',t)$, $p'({\bf x}',t)$, ${\bf f}'({\bf x}')$ denote the velocity and pressure of fluid, and the external force  in the coordinate system $O'-x'_{1}x'_{2}x'_{3}$, respectively, where ${\bf x}'\in[0,2\pi]^3$.  Certainly, they  also satisfy (\ref{eq:u}) and (\ref{eq:p})-(\ref{eq:p:m}) for $t\geq 0$, thus 
\begin{eqnarray}
{\bf u}'^{(1)}
 +\left( {\bf u}'\cdot \nabla'\right) {\bf u}' &=&-\nabla' p' + \frac{1}{Re} \Delta' {\bf u}' + {\bf f}',  \label{eq:u:B} \\
(m+1){\bf u}'^{(m+1)} 
 +\sum_{i=0}^{m}\left( {\bf u}'^{(i)}\cdot \nabla'\right) {\bf u'^{(m-i)}} &=& -\nabla' p'^{(m)} + \frac{1}{Re} \Delta' {\bf u}'^{(m)} ,   \;\; m\geq 1,   \hspace{1.0cm}  \label{eq:u:B:m}
\end{eqnarray}
and 
\begin{eqnarray}
\Delta' p' &=& \nabla'\cdot {\bf f}' -\nabla'\cdot  \left[\left( {\bf u'}\cdot \nabla'\right) {\bf u'}\right],  \label{eq:p:B}   \\ 
\Delta' p'^{(m)} &=&-\sum_{i=0}^{m}\nabla'\cdot  \left[\left( {\bf u}'^{(i)}\cdot \nabla'\right) {\bf u}'^{(m-i)}\right], \;\; m\geq 1. \label{eq:p:m:B}    
\end{eqnarray}

In general case,  ${\bf u}'({\bf x}',t) \neq {\bf u}({\bf x},t)$.   However, if there exists ${\bf u}'({\bf x}',t) = {\bf u}({\bf x},t)$, then there exists the spatial symmetry under the coordinate translation $x'_{i}=x_{i}+\pi$ and/or the coordinate rotation $x'_{i}=-x_{i}$.  

\begin{description}
\item[Lemma A]  {\em Under the coordinate translation $x'_{i}=x_{i}+\pi$ and/or the coordinate rotation $x'_{i}=-x_{i}$, it holds}
\begin{equation}
\nabla' = \nabla, \;\; \Delta' = \Delta.   \label{Lemma:A}
\end{equation}  
\end{description}

\hspace{-0.75cm}Proof:  Let ${\bf e}_{i}, {\bf e}'_{i}$ denote the axis unit vector for $x_{i}, x'_{i}$, respectively, where $i=1,2,3$.    For the coordinate translation $x'_{i}=x_{i}+\pi$,  it holds 
\begin{equation}
{\bf e}_{i} = {\bf e}'_{i}, \;\;\; \frac{\partial }{\partial x_{i}} = \frac{\partial }{\partial x'_{i}} \frac{\partial x'_{i}}{\partial x_{i}} = \frac{\partial }{\partial x'_{i}}, \;\;\; i=1,2,3,
\end{equation}  
thus 
\begin{equation}
\nabla' =\sum_{i=1}^{3} {\bf e}'_{i} \;  \frac{\partial }{\partial x'_{i}} =\sum_{i=1}^{3} {\bf e}_{i} \;  \frac{\partial }{\partial x_{i}} = \nabla.
\end{equation}
Similarly, for the coordinate rotation $x'_{i}=-x_{i}$, where $i=1,2,3$, it holds 
\begin{equation}
{\bf e}_{i} = -{\bf e}'_{i}, \;\;\; \frac{\partial }{\partial x_{i}} = \frac{\partial }{\partial x'_{i}} \frac{\partial x'_{i}}{\partial x_{i}} = -\frac{\partial }{\partial x'_{i}},
\end{equation}  
thus,   
\begin{equation}
\nabla' =\sum_{i=1}^{3} {\bf e}'_{i} \;  \frac{\partial }{\partial x'_{i}} =\sum_{i=1}^{3} \left(-{\bf e}_{i}\right) \; \left(- \frac{\partial }{\partial x_{i}}\right) =\sum_{i=1}^{3} {\bf e}_{i} \;  \frac{\partial }{\partial x_{i}} = \nabla.
\end{equation}
So, under  $x'_{i}=x_{i}+\pi$ and/or   $x'_{i}=-x_{i}$, it holds 
\begin{equation}
\Delta' = \nabla' \cdot \nabla' = \nabla \cdot \nabla =\Delta. 
\end{equation}
This ends the proof of Lemma A.  

\begin{description}
\item[Lemma B]  {\em 
For the incompressible Navier-Stokes equations (\ref{eq:continuation}) and (\ref{eq:u}) under periodic boundary condition, if there exist the spatial symmetry ${\bf u}'({\bf x}',t_{0}) = {\bf u}({\bf x},t_{0})$ at $t=t_{0}$ and ${\bf f}'({\bf x}')= {\bf f}({\bf x})$, where  $x'_i = x_i+\pi$ and/or  $x'_i = -x_i$ $(i=1,2,3)$, and besides  there exists the temporal Taylor series  of ${\bf u}({\bf x},t)$ at $t=t_{0}$, then there exist the spatial symmetry  ${\bf u}'^{(m+1)}({\bf x}',t_{0}) = {\bf u}^{(m+1)}({\bf x},t_{0})$ and   $p'^{(m)}({\bf x}',t_{0}) = p^{(m)}({\bf x},t_{0})$ at $t=t_{0}$ for all integer $m\geq 0$. 
}
\end{description}

\hspace{-0.75cm}Proof.  Lemma~B can be proved by means of recursive method. 

\hspace{-0.75cm} (A) When $m=0$, (\ref{eq:p:B}) becomes at $t=t_{0}$ that 
\begin{equation}
\Delta p'=\nabla\cdot {\bf f}  -\nabla\cdot  \left[\left( {\bf u}\cdot \nabla\right) {\bf u}\right], \label{eq:p:B:2}    
\end{equation}
since $\Delta' =\Delta, \nabla' = \nabla$ according to Lemma~A, and besides there exist the spatial symmetry ${\bf u}'({\bf x}',t_0) = {\bf u}({\bf x},t_{0})$ and ${\bf f}'({\bf x}') = {\bf f}({\bf x})$.   Comparing (\ref{eq:p:B:2}) with (\ref{eq:p}), we have  $\Delta p =\Delta p'$, that gives 
$p'({\bf x}',t_{0}) = p({\bf x},t_{0})$, i.e.  $p'^{(0)}({\bf x}',t_{0}) = p^{(0)}({\bf x},t_{0})$.   
 
Similarly, (\ref{eq:u:B}) becomes at $t=t_{0}$ that
 \begin{equation}
 {\bf u}'^{(1)}
 +\left( {\bf u}\cdot \nabla\right) {\bf u} =-\nabla p + \frac{1}{Re} \Delta {\bf u} + {\bf f},  \label{eq:u:B:2}
 \end{equation}
 since $\Delta' =\Delta, \nabla' = \nabla$ according to Lemma~A and there exist the spatial symmetry ${\bf u}'({\bf x}',t_0) = {\bf u}({\bf x},t_{0})$, ${\bf f}'({\bf x}') = {\bf f}({\bf x})$ and $p'({\bf x}',t_{0}) = p({\bf x},t_{0})$ as mentioned  above.  Comparing  (\ref{eq:u:B:2}) to (\ref{eq:u}) gives ${\bf u}'^{(1)}({\bf x}',t_0) = {\bf u}^{(1)}({\bf x},t_{0})$.  
 
 Thus, Lemma~B holds when $m=0$.  

\hspace{-0.75cm}(B) Assume that Lemma~B holds when $m = n \geq 0$, say, there exists the spatial symmetry 
 \begin{equation}
 {\bf u}'^{(i+1)}({\bf x}',t_{0}) = {\bf u}^{(i+1)}({\bf x},t_{0}),  \;\;  p'^{(i)}({\bf x}',t_{0}) = p^{(i)}({\bf x},t_{0}), \;\; 0\leq i\leq n. \label{symmetry:n}
 \end{equation} 
  Then, (\ref{eq:p:m:B} ) becomes  at $t=t_{0}$ that 
 \begin{equation}
 \Delta p'^{(n+1)} =-\sum_{i=0}^{n+1}\nabla\cdot  \left[\left( {\bf u}^{(i)}\cdot \nabla\right) {\bf u}^{(n+1-i)}\right],  \label{eq:p:n}
 \end{equation}
 since $\Delta' =\Delta, \nabla' = \nabla$ according to Lemma~A, and there exist the spatial symmetry (\ref{symmetry:n}).  Comparing (\ref{eq:p:n}) to (\ref{eq:p:m}) gives  $p'^{(n+1)}({\bf x}',t_{0}) = p^{(n+1)}({\bf x},t_{0})$.   
 
 Then, according to the spatial symmetry (\ref{symmetry:n}), (\ref{eq:u:B:m}) (when $m=n+1$) becomes at $t=t_{0}$  that  
  \begin{equation}
 (n+2){\bf u}'^{(n+2)} 
 +\sum_{i=0}^{n+1}\left( {\bf u}^{(i)}\cdot \nabla\right) {\bf u}^{(n+1-i)} = -\nabla p^{(n+1)} + \frac{1}{Re} \Delta {\bf u}^{(n+1)},   \label{eq:u:n:B}
 \end{equation}
 since $\Delta' =\Delta, \nabla' = \nabla$ according to Lemma~A and  $p'^{(n+1)}({\bf x}',t_{0}) = p^{(n+1)}({\bf x},t_{0})$ as  proved above.   Setting $m=n+1$ in (\ref{eq:u:m}) gives at $t=t_{0}$ that
\begin{equation}
(n+2){\bf u}^{(n+2)} 
 +\sum_{i=0}^{n+1}\left( {\bf u}^{(i)}\cdot \nabla\right) {\bf u}^{(n+1-i)} =-\nabla p^{(n+1)} + \frac{1}{Re} \Delta {\bf u}^{(n+1)}.  \;\;  \label{eq:u:n}
\end{equation}
Comparing (\ref{eq:u:n:B}) to (\ref{eq:u:n}) gives ${\bf u}'^{(n+2)}({\bf x}',t_{0}) ={\bf u}^{(n+2)}({\bf x},t_{0})$.  

Thus, if Lemma~B holds for $m=n\geq 0$, it holds for $m=n+1$.   
 
 \hspace{-0.75cm}(C) According to (A) and (B), Lemma~B holds for arbitrary integer $m\geq 0$.
 
 This ends the proof of Lemma~B.
 
 \begin{description}
\item[ Lemma C]  {\em For the incompressible Navier-Stokes equations (\ref{eq:continuation}) and (\ref{eq:u}) under periodic boundary condition, if there exist the spatial symmetry ${\bf u}'({\bf x}',t_{0}) = {\bf u}({\bf x},t_{0})$ at $t=t_{0}$ and ${\bf f}'({\bf x}')= {\bf f}({\bf x})$, where  $x'_i = x_i+\pi$ and/or  $x'_i = -x_i$ $(i=1,2,3)$,   and besides  there exists  the temporal Taylor series  of ${\bf u}({\bf x},t)$ at $t=t_{0}$ with a non-zero radius $\rho$ of convergence, then there exists the spatial symmetry  ${\bf u}'({\bf x}',t) = {\bf u}({\bf x},t)$ in $t\in[t_{0},\rho)$. }
\end{description}

\hspace{-0.75cm}Proof.  According to the Lemma~B, the temporal Taylor series of ${\bf u}({\bf x},t)$ at $t=t_{0}$ reads
\begin{eqnarray}
{\bf u}({\bf x},t) &=& \sum_{m=0}^{+\infty}  {\bf u}^{(m)}({\bf x},t_{0}) (t-t_{0})^{m}= \sum_{m=0}^{+\infty}  {\bf u}'^{(m)}({\bf x}',t_{0}) (t-t_{0})^{m}={\bf u}'({\bf x}',t) \hspace{1.0cm}
\end{eqnarray}
in $|t-t_{0}|<\rho_{0}$, which includes $t\in[t_{0},t_{0}+\rho_{0})$.  This ends the proof of Lemma~C.

\begin{description}
\item[Theorem of Spatial Symmetry Invariance of 3D NS equations] {\em 
The solution $\bf u$  of the incompressible Navier-Stokes equations (\ref{eq:continuation}) and (\ref{eq:u}) under periodic boundary condition has the same spatial symmetry ${\bf u}'({\bf x}',t)$ $ =  {\bf u}({\bf x},t)$ for all $t>0$ as its smooth initial condition, where $x'_i = x_i+\pi$ for translation  and/or   $x'_i = -x_i$ $(i=1,2,3)$ for rotation, if there exist the spatial symmetry ${\bf u}'({\bf x}',0)= {\bf u}({\bf x},0)$ for initial condition  and  ${\bf f}'({\bf x}')= {\bf f}({\bf x})$ for external force, and besides the temporal Taylor series of ${\bf u}({\bf x},t)$ exists and has a non-zero radius of convergence for $t\geq 0$. 
}
\end{description}

\hspace{-0.75cm}Proof.  Since there exists the spatial symmetry  ${\bf u}'({\bf x}',0) = {\bf u}({\bf x},0)$ and ${\bf f}'({\bf x}')= {\bf f}({\bf x})$, where $x'_i = x_i+\pi$ and/or  $x'_i = -x_i$  $(i=1,2,3$),  we have according to Lemma~C  (setting $t_{0}=0$) the spatial symmetry
\[  {\bf u}'({\bf x}',t) = {\bf u}({\bf x},t), \hspace{1.0cm} t\in[0,\delta_{0}],  \]
where $\delta_{0} <\rho_{0}$, and $\rho_{0}>0$ is the non-zero radius of convergence of its Taylor series at $t=0$.  

Similarly, since  there exists the spatial symmetry  ${\bf u}'({\bf x}',\delta_{0}) = {\bf u}({\bf x},\delta_{0})$ at $t=\delta_{0}$ and ${\bf f}'({\bf x}')= {\bf f}({\bf x})$, we have according to Lemma~C  (setting $t_{0}=\delta_{0}$) the spatial symmetry
\[  {\bf u}'({\bf x}',t) = {\bf u}({\bf x},t), \hspace{1.0cm} t\in[\delta_{0},\delta_{1}],  \]
where $\delta_{1} <\rho_{1}$, and $\rho_{1}>0$ is the non-zero radius of convergence of its Taylor series at $t=\delta_{0}$.

Note that 
\[   [0,\delta_{0}] \cup [\delta_{0},\delta_{1}] = [0,\delta_{1}],  \]
thus in this way there exists the spatial symmetry 
\[ {\bf u}'({\bf x}',t) = {\bf u}({\bf x},t), \hspace{1.0cm} t\in[0,\delta_{1}],  \]
in a larger interval of time.  The same process can keep going and the time interval becomes larger and larger, since the temporal Taylor series of ${\bf u}({\bf x},t)$ always exists with a non-zero radius of convergence.  
Thus,  it holds the spatial symmetry 
\[  {\bf u}'({\bf x}',t) = {\bf u}({\bf x},t), \hspace{1.0cm} t\in[0,+\infty). \] 
This ends the proof of the Theorem of Spatial Symmetry Invariance.  

\begin{description}
\item[Remark A]  The Theorem of Spatial Symmetry Invariance for Navier-Stokes equations can be used as a criterion to verify the correctness and accuracy of numerical simulation of Navier-Stokes equations in the corresponding conditions:  numerical simulations of  Navier-Stokes equations must depart from exact solution greatly if they distinctly violate this theorem.        
 \end{description}

 \begin{description}
\item[Remark B]     Let $O$ denote one observer in the coordinate system $O-x_{1}x_{2}x_{3}$, and $O'$ denote another observer in the coordinate system $O'-x'_{1}x'_{2}x'_{3}$, respectively, where $x'_{i}=x'_{i}+\pi$ and/or  $x'_{i}=-x'_{i}$.   When there exist the spatial symmetry  ${\bf f}'({\bf x}')= {\bf f}({\bf x})$ for external force and ${\bf u}'({\bf x}',0) = {\bf u}({\bf x},0)$ for initial condition,  the governing equations and the initial condition are the {\em same} for both of the observer $O$ in $O-x_{1}x_{2}x_{3}$ and the observer $O'$ in the  coordinate system  $O'-x'_{1}x'_{2}x'_{3}$.  Physically speaking, for the two observers at the two different coordinate systems, they should observe the same flow if the initial condition and the external force are the same, say,  ${\bf u}'({\bf x}',t)= {\bf u}({\bf x},t)$, for $t\in[0,+\infty)$.  From the viewpoint of the third observer knowing their results simultaneously, the flow governed by NS equations should have a kind of spatial symmetry.  This physical insight is indeed true as mathematically proved above in this paper.  So,   the Theorem of Spatial Symmetry Invariance is easy to understand from physical viewpoint.     
\end{description}

\section*{Acknowledgements}
This work is partly supported by National Natural Science Foundation of China (No. 12521002). 
	
\section*{Data and materials availability}
All data are available by sending requirement to the corresponding author.

\section*{Competing interests}
The authors declare no competing interests.

\bibliographystyle{elsarticle-num}
\bibliography{Kolmogorov}

\end{document}